\documentclass[epj,twocolumns]{svjour_mod}
\usepackage[utf8]{inputenc}
\usepackage{latexsym}
\usepackage{graphics}
\usepackage{amssymb}
\usepackage[colorlinks,citecolor=blue,urlcolor=blue,linkcolor=blue]{hyperref}
\usepackage{hyphenat}
\usepackage{amsmath}
\usepackage{cite}
\usepackage{relsize}
\usepackage{trimclip}
\usepackage{booktabs} 
\usepackage{tabularx}
\usepackage{multirow}
\usepackage{fixmath}
\usepackage{tikz}
\usetikzlibrary{calc}
\usepackage{placeins}
\usepackage{xspace}
\usepackage{slashed}
\newcommand{\order}[1]{{\cal O}\!\left(#1\right)}

\usepackage{color}

\definecolor{semiblue}{rgb}{0.3,0.3,0.8}
\newcommand{\logbook}[2]{}

\def\as{\alpha_{s}}

\def\d{\hbox{d}}

\makeatletter
\g@addto@macro\bfseries{\boldmath}
\makeatother

\begin{document}

\title{$C$-parameter hadronisation in the symmetric 3-jet
    limit and impact on $\alpha_s$ fits}
  
\newcommand{\CERNAff}{CERN, Theoretical Physics Department, CH-1211 Geneva 23, Switzerland }

\author{Gionata Luisoni\inst{1,}\thanks{Current address: MTU Aero
    Engines AG Dachauer Str. 665 80995 Munich, Germany} 
  \and Pier Francesco Monni\inst{1} 
  \and Gavin P. Salam\inst{2,3}}

\authorrunning{G.~Luisoni, P.~F.~Monni, G.~P.~Salam}

\institute{
\CERNAff
\and Rudolf Peierls Centre for Theoretical Physics, Clarendon Laboratory, Parks Road, University of
Oxford, Oxford OX1
3PU, UK
\and
All Souls College, Oxford OX1 4AL, UK
}

\date{}

\abstract{
  Hadronisation corrections are crucial in extractions of the strong
  coupling constant ($\alpha_s$) from event-shape distributions at
  lepton colliders.
  Although their dynamics cannot be understood rigorously using
  perturbative methods, their dominant effect on physical observables
  can be estimated in singular configurations sensitive to the
  emission of soft radiation.
  The differential distributions of some event-shape variables,
  notably the $C$ parameter, feature two such singular points.
  We analytically compute the leading non-perturbative correction in
  the symmetric three-jet limit for the $C$ parameter, and find that
  it differs by more than a factor of two from the known result in the
  two-jet limit.
  We estimate the impact of this result on strong coupling
  extractions, considering a range of functions to interpolate the
  hadronisation correction in the region between the 2 and 3-jet
  limits.
  Fitting data from ALEPH and JADE, we find that most interpolation
  choices increase the extracted $\as$, with effects of up to $4\%$
  relative to standard fits. 
  This brings a new perspective on the
  long-standing discrepancy between certain event-shape $\alpha_s$
  fits and the world average.
}

\PACS{{12.38.-t}{Quantum Chromodynamics}} 
\maketitle

\section{Introduction}
The strong coupling constant $\alpha_s$ is the least well known
coupling in the gauge sector of the Standard Model.
The latest Particle Data Group (PDG) average of $\alpha_s$ has an uncertainty of about
1\%~\cite{Tanabashi:2018oca,PDGQCD2019}, considerably larger than the
error in the other gauge coupling determinations.
Given the importance of QCD at LHC collider experiments, and
rapid progress in perturbative calculations~\cite{Ellis:2019qre} and
experimental accuracy, the uncertainty on $\as$ is becoming increasingly
critical for precision collider phenomenology.
However, the headline figure of 1\% uncertainty from the PDG average
masks significant discrepancies between different extractions.
In particular two of the most precise $\as$ determinations come from
event-shapes studies: $\as = 0.1135\pm0.0010$~\cite{Abbate:2010xh}
from fitting thrust data and
$\as = 0.1123\pm0.0015$~\cite{Hoang:2015hka} from $C$-parameter
data.
These results are several standard deviations away from the world
average of $0.1179 \pm0.0010$~\cite{Zyla:2020zbs,PDGQCD2019} and
from other individual precise extractions, such as $0.1185\pm 0.0008$
from lattice step scaling~\cite{Bruno:2017gxd} and $0.1188\pm 0.0013$
from jet rates \cite{Verbytskyi:2019zhh}.

These particular event-shape and jet-rate fits are among the most
precise of a wide variety of fits to $e^+e^-$ hadronic final-state
data~\cite{Jones:2003yv,Dissertori:2007xa,Bethke:2008hf,Becher:2008cf,Davison:2008vx,Dissertori:2009ik,Gehrmann:2009eh,Chien:2010kc,Abbate:2010xh,OPAL:2011aa,Gehrmann:2012sc,Abbate:2012jh,Hoang:2015hka,Kardos:2018kqj,Dissertori:2009qa,Schieck:2012mp,Verbytskyi:2019zhh,Kardos:2020igb}.
Many of them use high-precision perturbative calculations,
however they also all require input on non-perturbative
(hadronisation) effects.
These can be estimated either using Monte Carlo (MC) event
generators~\cite{Bethke:2008hf,Dissertori:2009ik,OPAL:2011aa,Kardos:2018kqj,Verbytskyi:2019zhh}
or via analytic non-perturbative
models~\cite{Gardi:2003iv,Davison:2008vx,Abbate:2010xh,Gehrmann:2012sc,Hoang:2015hka}.
The use of MC event generators has long been criticised on two main
grounds: they are tuned on less accurate perturbative (shower)
calculations, and the separation between perturbative and
non-perturbative components cannot easily be related to today's
highest-accuracy perturbative calculations.
Conversely the analytic models fit a non-perturbative parameter and
the perturbative coupling in a single, consistent framework.
The low $\as$ values from
Refs.~\cite{Abbate:2010xh,Gehrmann:2012sc,Hoang:2015hka} use the
latter method.
The price to pay in this approach is that the non-perturbative
component is not controlled beyond the first order in an expansion in
powers of $1/Q$ (the centre-of-mass energy) and furthermore only in
the $2$-jet limit, while fits cover both the $2$ and $3$-jet regions.

In this article we examine specifically the issue of going beyond the
$2$-jet limit for the hadronisation correction.
In principle one might attempt a full calculation as carried out for
top-quark production in the large-$n_f$ limit in
Ref.~\cite{FerrarioRavasio:2018ubr}.
Before embarking on such a calculation, we believe however that it is
worth establishing whether $3$-jet hadronisation corrections bring a
phenomenologically relevant effect.
To do so simply, we consider the example of the $C$-parameter.
This observable is special in that it has two singular points, one at
$C=0$ and the other at a Sudakov shoulder at
$C=3/4$~\cite{Catani:1997xc,Catani:1998sf}.
Existing fits calculate the hadronisation correction around the first
singular point, $C=0$ and extend it to the whole $C$-parameter spectrum.
Here we point out that one can also quite straightforwardly calculate
the power correction at the other singular point $C=3/4$.
One can then consider a range of schemes for interpolating between the
two singular points and examine their impact on strong coupling fits. 

This letter is structured as follows: in
Section~\ref{sec:perturbative} we briefly recall the framework for
perturbative fits with analytic hadronisation estimates.
In Section~\ref{sec:non-perturbative} we then review the determination
of the $C$-parameter hadronisation correction in the $2$-jet limit and
extend it to the symmetric $3$-jet case ($C=3/4$).
Section~\ref{sec:fit} presents the results of our new fits and we then
conclude in Section~\ref{sec:conclusions}.

\section{The $C$-parameter and its distribution}
\label{sec:perturbative}
The $C$-parameter variable for a hadronic final state in $e^+e^-$
annihilation is defined as follows~\cite{Ellis:1980wv},
\begin{equation}
  \label{eq:C-from-eigenvalues}
  C = 3 \left(\lambda_1\lambda_2 + \lambda_2\lambda_3 + \lambda_3\lambda_1\right),
\end{equation}
in terms of the eigenvalues $\lambda_i$ of the linearised momentum
tensor $\Theta^{\alpha\beta}$ \cite{Parisi:1978eg,Donoghue:1979vi},
\begin{equation}
  \label{eq:momentum tensor}
  \Theta^{\alpha\beta} = \frac1{\sum_i |\vec p_i|}
  \sum_i \frac{{\vec p}_i^\alpha {\vec p}_i^\beta}{|\vec p_i|}\,,
\end{equation}
where $|\vec p_i|$ is the modulus of the three momentum of particle
$i$ and ${\vec p}_i^\alpha$ is its momentum component along 
spatial dimension $\alpha$ ($\alpha = 1,2,3$).
In events where all particles are massless, this can also be written
as
\begin{align}
C &= 3 - \frac{3}{2}\sum_{i,j} \frac{(p_i\cdot p_j)^2}{(p_i\cdot Q)
    (p_j\cdot Q)}=\frac{3}{8}\sum_{i,j} x_i x_j \sin^2\theta_{ij}\,,
\label{eq:Cpar}
\end{align}
where $Q$ is the centre-of-mass energy, $p_i$ denotes the
four-momentum of particle $i$, $x_i = 2 (p_i\cdot Q)/Q^2$, and
$\theta_{ij}$ is the angle between particles $i$ and $j$.
We introduce the cumulative distribution ${\rm \Sigma}$ defined as
\begin{equation}
\label{eq:sigma}
  {\rm \Sigma}(C)\equiv \frac{1}{\sigma}\int_0^{C}\d C' 
  \frac{\d\sigma}{\d C'}\,.
\end{equation}
The differential distribution $\d\sigma$ is known to
next-to-next-to-leading order (NNLO) in massless
QCD~\cite{GehrmannDeRidder:2007hr,Weinzierl:2009ms,DelDuca:2016csb},
which can be combined with the total cross section
$\sigma$~\cite{Gorishnii:1990vf} to obtain a N$^3$LO prediction for
Eq.~\eqref{eq:sigma}.
The effects of heavy-quark (notably the bottom quark) masses on event
shape distributions~\cite{Nason:1997nw}, as well as electroweak
corrections~\cite{Denner:2009gx}, are known to NLO, but we do not
consider them in our study. Their omission does not affect in any way
the conclusions of this article. In the following we consider the
massless-QCD NNLO predictions for the differential distribution from
Ref.~\cite{DelDuca:2016csb}.

In the two-jet region, the fixed-order expansion is spoiled by large
logarithms of infrared and collinear origin, which must be consistently
resummed at all perturbative orders to obtain a physical prediction.
The resummation for the $C$-parameter distribution has been carried
out in different
formalisms~\cite{Catani:1998sf,Hoang:2014wka,Banfi:2014sua} and it is
known up to N$^3$LL~\cite{Hoang:2014wka}. In our analysis we adopt the
analytic next-to-next-leading logarithmic (NNLL) calculation from the appendix of
Ref.~\cite{Banfi:2014sua}, which is sufficient to illustrate our
findings.

To obtain a perturbative prediction that is accurate across the whole
physical spectrum, we need to match the resummed NNLL calculation to
the fixed order result. This is done by combining the N$^3$LO
calculation ${\rm \Sigma}^{\rm N^3LO}(C)$ with the resummed prediction
${\rm \Sigma}^{\rm NNLL}(C)$ according to the log-R
scheme~\cite{Catani:1992ua} as
\begin{equation}
  \Sigma^\text{pert.}(C) \equiv e^{\ln {\rm \Sigma}^{\rm NNLL}(C) + \ln
    {\rm \Sigma}^{\rm N^3LO}(C) - \ln{\rm \Sigma}^{\rm Exp.}(C)}\,,
\end{equation}
where ${\rm \Sigma}^{\rm Exp.}(C)$ is the fixed-order expansion of
${\rm \Sigma}^{\rm NNLL}(C)$ to ${\cal O}(\alpha_s^3)$.
Detailed formulae are reported in Ref.~\cite{Monni:2011gb}.
Note that specific choices need to be made to limit the impact of the
resummation in regions where $C$ is not small.
Our choices are discussed in Appendix~\ref{app:modlogs}.

The hadronisation corrections to the $C$ parameter distribution can be
described in terms of an expansion in negative powers of the centre of
mass energy $Q$. The leading correction in this sense leads to a shift
of the cumulative distribution of the form
\begin{equation}
  \Sigma^\text{hadr.}(C) = \Sigma^\text{pert.}(C - \langle \delta C\rangle(C))\,,
\end{equation}
where $\langle \delta C\rangle(C) \propto 1/Q$ is the mean change in the
$C$ parameter's value due to the emission of soft non-perturbative
radiation.
In most work, the power correction is taken to be independent of the
value of the observable, $\langle \delta C\rangle(C) \equiv \langle
\delta C\rangle(0)$.
In this work we will be investigating the consequences of having
$\langle \delta C\rangle(C)$ vary with $C$.

We adopt the following form for $\langle \delta C\rangle(C)$,
\begin{align}
\label{eq:deltaC-master}
  \langle \delta C\rangle &\simeq  \zeta(C)\,{\cal M}\,\frac{\mu_{I}}{Q}\frac{4 \,C_F}{\pi^2} \bigg[\alpha_{0}(\mu_{I}^{2})-\alpha_{s}(\mu_{R}^2)\nonumber\\
                          &-\alpha_{s}^{2}(\mu_{R}^2)\frac{\beta_{0}}{\pi}\left(2\ln\frac{\mu_{R}}{\mu_{I}}+\frac{K^{(1)}}{2\beta_{0}}+2\right)\nonumber\\
                          &-\alpha_{s}^{3}(\mu_{R}^2)\frac{\beta_{0}^2}{\pi^2}\bigg(4\ln^{2}\frac{\mu_{R}}{\mu_{I}}+4\bigg(\ln\frac{\mu_{R}}{\mu_{I}}+1\bigg)\nonumber\\
                          &\times\bigg(2+\frac{\beta_{1}}{2\beta_{0}^2}+\frac{K^{(1)}}{2\beta_{0}}\bigg)+\frac{K^{(2)}}{4\beta_{0}^2}\bigg)\bigg]\,,
\end{align}
where $\zeta(C)$ encodes the $C$-dependence of the correction.
In Eq.~\eqref{eq:deltaC-master} we include the Milan factor
${\cal M}(n_f=3) \simeq
1.490$~\cite{Dokshitzer:1997iz,Dasgupta:1999mb,Beneke:1997sr} to
account for the non-inclusive correction.
Eq.~(\ref{eq:deltaC-master}) involves the mean value of the strong
coupling constant in the soft physical scheme $\tilde{\alpha}_s$ (see
also Appendix~\ref{app:formulae}) at infrared scales $\mu \leq \mu_I$,
above which the prediction is assumed to be dominantly
perturbative~\cite{Dokshitzer:1995qm},
\begin{align}
  \alpha_0(\mu_I^2) \equiv \frac{1}{\mu_I}\int_0^{\mu_I}\d \mu\,\tilde{\alpha}_s(\mu^2)\,.
\end{align}
Eq.~(\ref{eq:deltaC-master}) also includes terms to subtract the
contributions already accounted for in the perturbative
calculation~\cite{Dokshitzer:1995zt,Davison:2008vx,Gehrmann:2012sc}. 
The determination of the latter is not without subtleties, in that it
assumes that non-inclusive corrections to such renormalon subtraction
are described by the same multiplicative ${\cal M}$ factor as for the
coefficient of $\alpha_0(\mu_I^2)$.
However these subtleties are numerically subdominant relative to other
effects that we will be discussing here.

The core of this article relates to the coefficient $\zeta(C)$, which
entirely determines the $C$ dependence of
Eq.~(\ref{eq:deltaC-master}).
The calculation of $\zeta(C)$ at specific values of $C$ is the subject of the
next section.

\section{Non-perturbative corrections}
\label{sec:non-perturbative}
The calculation of $\langle \delta C\rangle$ near a singular
configuration requires the amplitudes describing the emission of a
``gluer''~\cite{Dokshitzer:1995qm} $k$ off the hard partonic system
defined by the set of momenta $\{p_i\}$.
We can then calculate the mean change in the observable
caused by the emission of $k$, i.e.
\begin{equation}
\label{eq:deltaC-def}
\Delta C (\{p'_i\},\{p_i\}; k) \equiv C (\{p'_i\}; k) - C(\{p_i\})\,.
\end{equation}
where the momenta $\{p_i\}$ ($\{p'_i\}$) describe the hard
configuration before (after) the emission of the gluer.
One can construct a range of prescriptions for mapping
$\{p_i\} \to \{p'_i\}$ and in general $\Delta C (\{p'_i\},\{p_i\}; k)$
depends on the choice that is made.
However, in the immediate vicinity of a singular configuration, it
turns out that $\Delta C (\{p'_i\},\{p_i\}; k)/k_t$ is independent of
that prescription in the limit of $k_t \to 0$, where
$k_t$ is the transverse momentum of $k$.
For the $C$-parameter, the singular points correspond to the 2-jet
limit and the symmetric 3-jet limit.
At these points the dependence on the choice of recoil scales as
$k_t^2$ and so vanishes in the $k_t \to 0$ limit of $\Delta C
(\{p'_i\},\{p_i\}; k)/k_t$. 

Under such conditions, one can write
\begin{multline}
  \label{eq:zetaC-def}
  \zeta(C) =
  \lim_{\epsilon\to0}
  \frac{\pi Q}{2\as C_F}
  \int [dk] \, M^2(k)\,
  \times \\ \times
  \Delta C (\{p'_i\},\{p_i\}; k)\,
  \delta(k_t - \epsilon)\,,
\end{multline}
where $[\d k] M^2(k)$ is the phase space and eikonal squared amplitude
describing the emission of the gluer, and the $\{p_i\}$ are the hard
momenta associated with the singular configuration.
The coupling, colour factor and a dimensional factor $Q$ are divided
out, since these are included in Eq.~(\ref{eq:deltaC-master}).

For an emission from a dipole $\{ij\}$, stretching between particles with
momenta $p_i$ and $p_j$, the matrix element and phase space are
\begin{equation}
  \label{eq:phasespace-ME}
  [dk] \, M^2(k) =
  \frac{2C_F}{\pi} \as \frac{dk_t}{k_t} \frac{d\phi}{2\pi} d\eta\,,
\end{equation}
where $k_t$, $\eta$ and $\phi$ are to be understood with respect to
the dipole;
$k_t$ in the $\delta(k_t-\epsilon)$ factor in Eq.~(\ref{eq:zetaC-def})
is also to be understood with respect to the emitting dipole.

\subsection{Calculation of $\zeta(0)$ (2-jet limit)}
\label{sec:C-2jet}

Let us start by considering the two-jet limit $C=0$.
The shift in the $C$-parameter induced by a small-$k_t$ gluer
is~\cite{Catani:1992ua,Catani:1998sf}
\begin{equation}
  \label{eq:deltaC-2jet-result}
  \Delta C(k) =  \frac{k_t}{Q} \frac{3}{\cosh(\eta)} +
  {\cal O}\left(\frac{k_t^2}{Q^2}\right)\,,
\end{equation}
where, for brevity, we have omitted the $\{p\}$ and $\{p'\}$
arguments in $\Delta C(k)$. 
This leads us to
\begin{equation}
  \label{eq:zeta0}
  \zeta(0) = \int_{-\infty}^{\infty}\d \eta\,
  \frac{3}{\cosh(\eta)}= 3 \pi \simeq 9.42478 \,,
\end{equation}
where the rapidity limits can be taken to infinity because the
integral is convergent.
This coincides (to within conventions for normalisations) with the
result that was given in
\cite{Dokshitzer:1995zt,Catani:1998sf,Dokshitzer:1998pt}.

\subsection{Calculation of $\zeta(3/4)$ (Sudakov shoulder)}
\label{sec:C-3jet}

The leading order (LO) $C$-parameter distribution has an endpoint at
$3/4$. 
Just below this endpoint, the distribution tends to a non-zero constant~\cite{Catani:1997xc},
\begin{equation}
\frac{1}{\sigma}\frac{\d\sigma}{\d C}\left( \frac{3}{4}^- \right) =
\frac{\alpha_s}{2\pi} C_F\,\frac{256}{243}\pi\sqrt{3} + {\cal O}(\alpha_s^2)\,,
\end{equation}
while above the endpoint the distribution is zero at order $\as$.
This structure is known as a Sudakov shoulder.
Parametrising the energies of the two quarks and the gluon as
$Q/2 (2/3 - \epsilon_q)$, $Q/2 (2/3 - \epsilon_{\bar q})$,
$Q/2 (2/3 + \epsilon_q + \epsilon_{\bar q})$
one obtains
\begin{equation}
  \label{eq:C-near-34}
  C = \frac{3}{4} - \frac{81}{16}\left(\epsilon_q^2 +
    \epsilon_q\epsilon_{\bar q} + \epsilon_{\bar q}^2 \right)
  + \order{\epsilon^3}
  \,,
\end{equation}
and the shoulder arises because of the absence of linear dependence on
the $\epsilon$'s.
This absence of linear dependence on $\epsilon$ is also the reason that the
choice of recoil prescription affects $\Delta C$ only at order
$k_t^2/Q^2$.
Considering emission of a gluon with momentum $k$ from an $\{i j\}$ dipole (with
$i,j$ chosen among $q,\bar{q},g$), one can then derive
\begin{equation}
  \label{eq:DeltaC-34}
  \Delta C(k) = \frac{3\sqrt{3}}{2}\,\frac{\sin^2(\phi)}{2 \cosh(\eta) -
    \cos(\phi)} \,\frac{ k_t}{Q} + {\cal O}\left(\frac{k_t^2}{Q^2}\right)\,,
\end{equation}
in terms of the $k_t$, $\eta$ and $\phi$ of the emission with respect to
the dipole (taken in the dipole's centre of mass).
The $\order{k_t/Q}$ contribution arises only when $k$ is out of the
3-jet plane.

The squared matrix element times phase space can be written as a sum
over three dipoles
\begin{equation}
  \label{eq:3jet-ME-sum}
  [dk]M^2(k) =
  \frac{2\as}{\pi}
  \!\!\!
  \sum_{\text{dip}=qg,\bar qg,q\bar q}
  \!\!\!\!
  C_\text{dip} \frac{dk_t^{[\text{dip}]}}{k_t^{[\text{dip}]}} \frac{d\phi^{[\text{dip}]}}{2\pi} d\eta^{[\text{dip}]},
\end{equation}
where $C_{qg} = C_{\bar q g} = C_A/2$ and $C_{q\bar q} = C_F - C_A/2$.
Note that for each dipole we will use the corresponding 
kinematic variables ($k_t^\text{[dip]}$, etc.) in
Eq.~(\ref{eq:DeltaC-34}).
This is equivalent to the procedure used to calculate the power
correction to the $D$-parameter for arbitrary $3$-jet
configurations~\cite{Banfi:2001pb}. 

Integrating over $\eta^\text{[dip]}$ and $\phi^\text{[dip]}$, and
summing over dipoles, we then obtain
\begin{align}
\zeta(3/4)& = \frac{3\sqrt{3}}{4}\frac{C_A+2 C_F}{C_F}\!\int_{-\infty}^{\infty}\!\!\!\!d\eta\int_{0}^{2\pi}\!\frac{d\phi}{2\pi}\frac{\sin^2\phi}{2\cosh\eta-\cos\phi}\notag\\
&=
\frac{3\sqrt{3}}{4}\, \frac{C_A + 2 C_F}{C_F}\left( 4\, E(1/4) -3\, K(1/4)\right)\,.
\end{align}
The functions $K$ and $E$ are the complete elliptic integrals of the
first and second kind
\begin{subequations}
\begin{align}
K(t) &= \int_0^{\pi/2} d \theta \left(1-t
       \sin^2\theta\right)^{-1/2}\,,\\
E(t) &=  \int_0^{\pi/2} d \theta \left(1-t
       \sin^2\theta\right)^{1/2}\,.
\end{align}
\end{subequations}
The numerical value of $\zeta(3/4)$ reads
\begin{equation}
\zeta(3/4) \simeq 4.48628\,,
\end{equation}
which provides the leading non-perturbative correction at the
shoulder.\footnote{The numerical value of $\zeta(3/4)$ was previously
  estimated in unpublished work by one of us (GPS) in collaboration with
  Z. Tr\'ocs\'anyi (see for instance Section 4.1.3 of
  Ref.~\cite{Dasgupta:2003iq}).}
This simple result reveals that the leading ($\sim 1/Q$) hadronisation
correction at the (symmetric three-jet) Sudakov shoulder is less than
half that in the two-jet limit ($\zeta(0) = 3\pi$).

\subsection{Modelling of the $0\! <\! C\! <\! 3/4$ region}
\begin{figure}
 \includegraphics[width=0.96\linewidth]{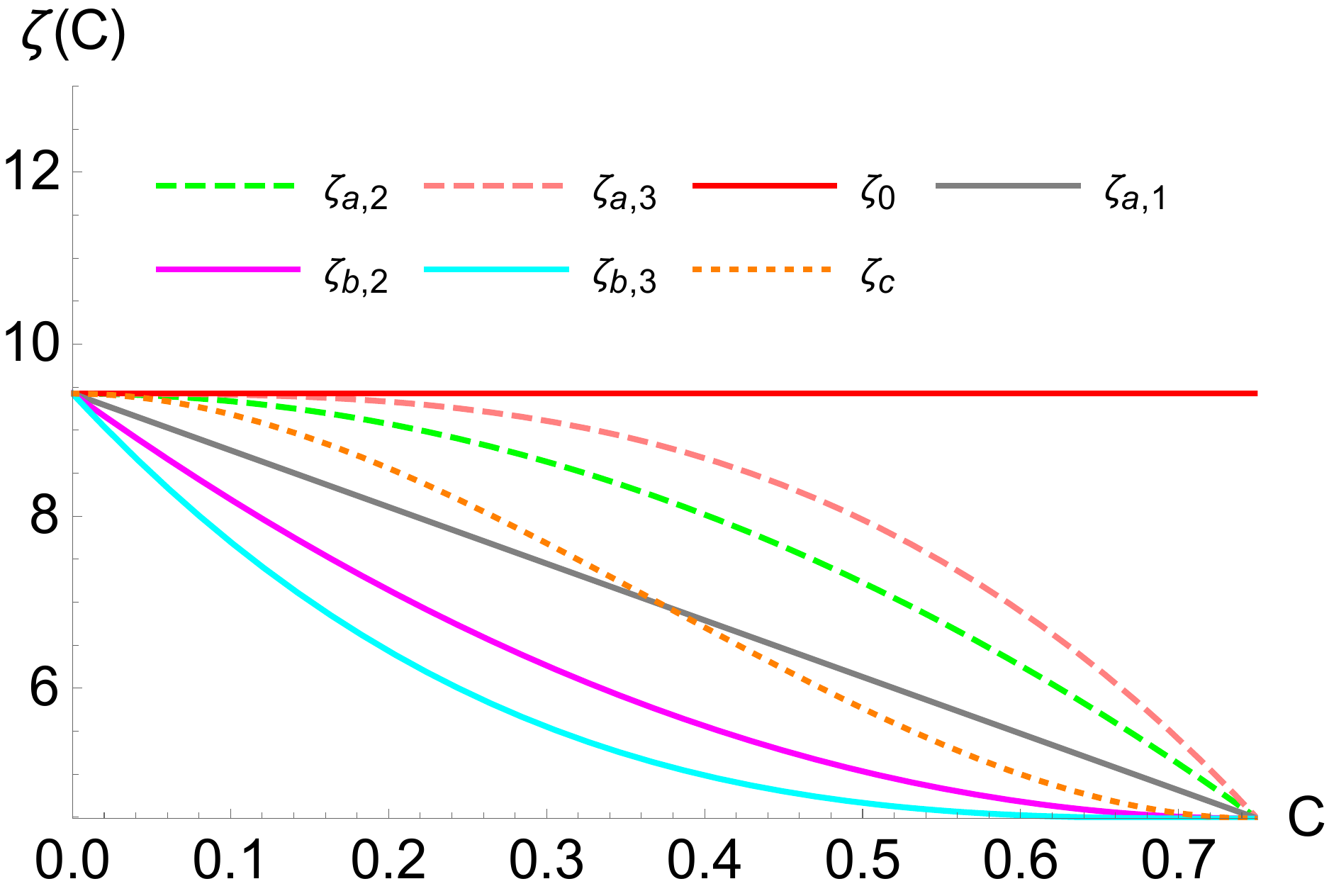}
 \caption{Different functional forms for $\zeta(C)$ function
   interpolating between the results at $C=0$ and $C=3/4$.}
 \label{fig:zetaC_profiles}
\end{figure}

Our calculations of $\zeta(0)$ and $\zeta(3/4)$ relied critically on
the fact that recoil from the gluer emission had an impact that was
quadratic in the gluer momentum.
Away from these special points, the methods used here do not give us
control over the value of the power correction, because the result
depends on the prescription that we adopt for recoil (the impact of
the hard parton's recoil becomes linear in the gluer momentum).
One could conceivably extend the methods of
Ref.~\cite{FerrarioRavasio:2018ubr} to attempt to determine the
general dependence of $\zeta(C)$ on $C$, however such a calculation is
highly non-trivial.
So here, we want to establish whether such a calculation would be
phenomenologically important.
To do so, we consider a range of models that interpolate
the power correction between the known values at $C=0$ and $C=3/4$,
some of which depend on a parameter $n \ge 0$. These are:
\begin{subequations}\label{eq:zeta-forms}
  \begin{align}
    \zeta_0(C)     &= \zeta(0)
                     \\
    \zeta_{a,n}(C) &= \zeta(0) (1-u^n) + \zeta(3/4)u^n\,,\quad
                     u = \frac{4C}{3}\,,
    \\
    \zeta_{b,n}(C) &= \zeta(0) (1-u)^n + \zeta(3/4)\left(1 - (1-u)^n\right),
    \\
    \zeta_c(C)     &= \zeta(0) + (\zeta(3/4)-\zeta(0)) g(u),
  \end{align}
  where $g(u)$ has the property that it is $0$ ($1$) for $u=0$ ($1$)
  and its first derivative is zero at $u=0,1$,
  \begin{equation}
    \label{eq:g_of_u}
    g(u) = -1 + (1-u)^3 + 3u -u^3\,.
  \end{equation}
\end{subequations}
The different forms for $\zeta(C)$ are shown in
Fig.~\ref{fig:zetaC_profiles}.
The $\zeta_0$ choice corresponds to using a constant shift, i.e.\ the
standard approach for earlier studies.
For both $\zeta_{a,n}$ and $\zeta_{b,n}$, using $n=1$ corresponds to a
linear interpolation between the $\zeta(0)$ and $\zeta(3/4)$ values.
For larger $n$, $\zeta_{a,n}$ is flat close to $C=0$, while
$\zeta_{b,n}$ is flat close to $C = 3/4$.
Finally $\zeta_{c}$ is flat near both $C=0$ and $C=3/4$.
We stress that the variations in Eqs.~(\ref{eq:zeta-forms}) are not
normally taken into account when estimating hadronisation with
analytic models, which effectively all assume the $\zeta_0$ model,
corresponding to a constant shift across the whole differential
distribution.
In Section~\ref{sec:fit} we will see what impact this has on fits for
the strong coupling from experimental data.

In order to gain some insight on how $\zeta(C)$ depends on the recoil
scheme, in Appendix~\ref{app:fixed-order} we carry out a fixed-order
calculation of this quantity within different schemes to distribute
the recoil due to the emission of the gluer among the remaining three
partons.
In reality, however, the behaviour that we find at fixed order in
Appendix~\ref{app:fixed-order} can be substantially modified by the
emission of multiple perturbative radiation (as also discussed in
Appendix~\ref{app:fixed-order}).
Therefore we do not rely on these
calculations to assess the impact of $\zeta(C)$ on the fits, but
rather use them as an insightful picture of how the leading
non-perturbative correction scales across the spectrum of the event
shape.
We do however note that the concrete recoil schemes all yield shapes
that fall below the $\zeta_{a,1}\equiv\zeta_{b,1}$ line.

\section{Fit of $\alpha_s$ and hadronisation uncertainties}
\label{sec:fit}
To test how our results affect the extraction of $\alpha_s$, we
perform a simultaneous fit of the strong coupling and of the
non-perturbative parameter $\alpha_0(\mu_I^2)$, using data at
different centre-of-mass energies from the {\tt ALEPH}~\cite{Heister:2003aj}
and {\tt JADE}~\cite{MovillaFernandez:1998ys} experiments, as summarised in
Table~\ref{tab:data}.
This dataset is smaller than that considered for a
similar fit in Ref.~\cite{Hoang:2015hka}, but is largely sufficient for
determining how the $\as$ fit result depends on $\zeta(C)$.
\begin{table}
\begin{tabular}{lrcrc}
  \toprule
  Exp. & Q (GeV) & Fit range  & N.~bins & Ref. \\
  \midrule
  {\tt ALEPH}      & 91.2    &  $0.27<C<0.69$  & 22 & \cite{Heister:2003aj} \\
  {\tt ALEPH}      & 133.0   &  $0.20<C<0.675$  &  6 & \cite{Heister:2003aj} \\
  {\tt ALEPH}      & 161.0   &  $0.16<C<0.675$  &  7 & \cite{Heister:2003aj} \\
  {\tt ALEPH}      & 172.0   &  $0.16<C<0.675$  &  7 & \cite{Heister:2003aj} \\
  {\tt ALEPH}      & 183.0   &  $0.16<C<0.675$  &  7 & \cite{Heister:2003aj} \\
  {\tt ALEPH}      & 189.0   &  $0.16<C<0.675$  &  7 & \cite{Heister:2003aj} \\
  {\tt ALEPH}      & 200.0   &  $0.125<C<0.675$  &  8 & \cite{Heister:2003aj} \\
  {\tt ALEPH}      & 206.0   &  $0.125<C<0.675$  &  8 & \cite{Heister:2003aj} \\
  {\tt JADE}       &  44.0   &  $0.61<C<0.68$  &  2 & \cite{MovillaFernandez:1998ys}    \\
  \bottomrule
\end{tabular}
\caption{Data set considered for the simultaneous $\chi^{2}$ fit
  of $\alpha_s$ and $\alpha_0$.}
    \label{tab:data}
\end{table}

The theory predictions are obtained using 50 bins in the
$0\leq C \leq 1$ range, subsequently interpolated in order to be
evaluated in correspondence to the experimental data bins.
The fit is performed by minimising the $\chi^2$ function defined as
\begin{align} 
\label{eq:chi2}
\chi^2 =& \sum_{i,j}
\left(\left.\frac{1}{\sigma}\frac{\d\sigma}{\d C}(C_i)\right|^{\rm
data}-\left.\frac{1}{\sigma}\frac{\d\sigma}{\d C}(C_i)\right|^{\rm
th}\right)V^{-1}_{ij}\notag\\
&\times\left(\left.\frac{1}{\sigma}\frac{\d\sigma}{\d C}(C_j)\right|^{\rm
data}-\left.\frac{1}{\sigma}\frac{\d\sigma}{\d C}(C_j)\right|^{\rm
th}\right)\, ,
\end{align}
where $V_{ij}$ is the covariance matrix that encodes the correlation
between the bins $C_i$ and $C_j$.  The general form of the covariance
matrix is $V_{ij}= S_{ij}+E_{ij}$, where
$S_{ij}=\delta\sigma_{{\rm stat},\,i}^2\delta_{ij}$ is the diagonal
matrix of the (uncorrelated) statistical errors in the experimental
differential distribution, while $E_{ij}$ contains the experimental
systematic covariances.  The diagonal entries of
$E_{ii}=\delta\sigma_{{\rm syst},i}^2$ are given by the experimental
systematic uncertainty on the $i$-th bin.
For the off-diagonal elements, which are not publicly available, a
common choice (used also in
Refs.~\cite{Abbate:2010xh,Gehrmann:2012sc,Hoang:2015hka}) is to
consider a
minimal-overlap model,
which defines $E_{ij}$ as
\begin{equation} 
  \label{eq:minoverlap}
  E_{ij} = {\rm min}\left(\delta\sigma_{{\rm
        syst},i}^2,\delta\sigma_{{\rm syst},j}^2\right).
\end{equation}
For ease of comparison, we adopt the same choice, though we note that
for the normalised distributions that we fit here, the true covariance
matrix would also include some degree of anti-correlation.
The $\chi^2$ minimisation is carried out with the {\tt TMinuit}
routine distributed with {\tt ROOT} and the whole analysis was
implemented in the {\tt C++} code used for a similar fit in
Ref.~\cite{Gehrmann:2012sc}.
Results with a diagonal covariance matrix, i.e.\ without any
correlations, are given in Appendix~\ref{app:uncorr}.
They yield almost identical central results for $\as$ and $\alpha_0$,
smaller $\chi^2$ values, and an increase in the experimental errors of
${\cal O}(10\%-20\%)$, which however remain small compared to
theoretical uncertainties.

In order to estimate the theoretical uncertainties, we perform the
following variations:
\begin{itemize}
\renewcommand{\labelitemi}{$\bullet$}
\item the renormalisation scale $\mu_R$ is randomly varied in the
  range $Q/2\leq \mu_R \leq 2\,Q$, while the infrared scale $\mu_I$ is
  set to $2$ GeV;
\item for $\mu_R = Q$, the resummation scale fraction $x_C$ defined in
  Appendix~\ref{app:modlogs} (default value $x_C=1/2$) is randomly
  varied by a factor $3/2$ in either direction, namely in the range
  $1/3 \leq x_C \leq 3/4$, following the prescription of
  Ref.~\cite{Jones:2003yv};
\item for $\mu_R = Q$ and $x_C = 1/2$, the Milan factor $\mathcal{M}$
  is randomly varied within $20\%$ of its central
  value~\cite{Dokshitzer:1997iz} (${\cal M}\simeq 1.49$) to account
  for non-inclusive effects in the $\langle \delta C\rangle$
  shift~\eqref{eq:deltaC-master} beyond ${\cal O}(\alpha_s^2)$;
\item keeping all of the above parameters at their central values,
  the parameter $p$ in the modified logarithm defined in
  Eq.~\eqref{eq:modlog} of Appendix~\ref{app:modlogs} (default value
  $p=6$) is replaced by $p=5$ and $p=7$. This choice for $p$ is
  discussed in Appendix~\ref{app:modlogs}.
\end{itemize}
The theory error is defined as the envelope of all the above
variations.
When we quote overall results below, we add the theoretical and
experimental errors in quadrature.

We test several models for $\zeta(C)$ as given in
Eq.~\eqref{eq:zeta-forms} and shown in
Fig.~\ref{fig:zetaC_profiles}.
Specifically, we consider the constant $\zeta_0$ choice, the
$\zeta_{a,n}$ model for $n=1,2,3$, the $\zeta_{b,n}$ model for
$n=1,2,3$, and the $\zeta_{c}$ model (recall $\zeta_{a,1} \equiv
\zeta_{b,1}$). 

\begin{figure}
 \includegraphics[width=\linewidth]{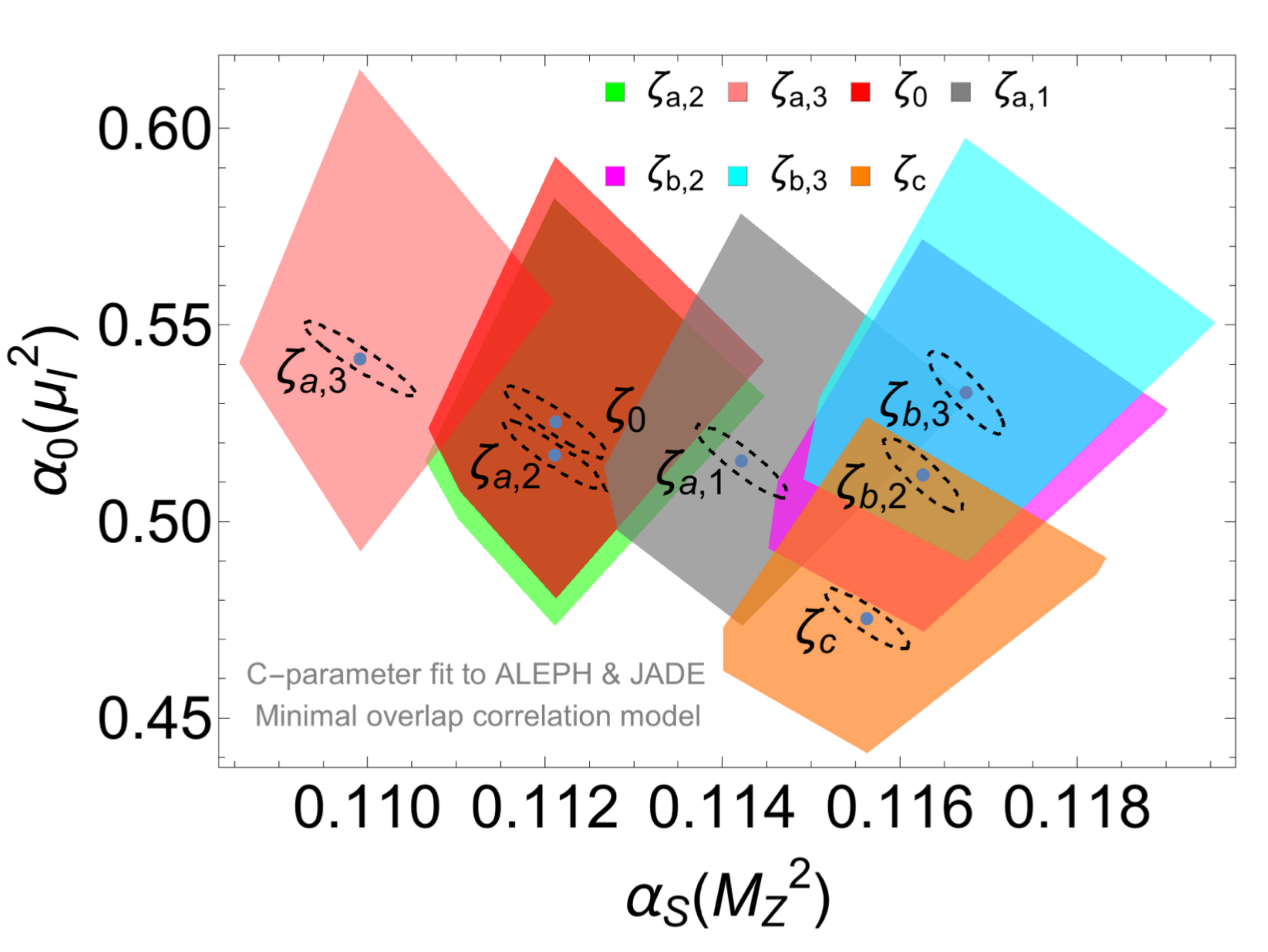}
 \caption{Fit results for $\alpha_s$ and $\alpha_0$ for different
   models of $\zeta(C)$.
   The points
   indicate the fit corresponding to the central setup of scales and
   parameters for a given model.
   The ellipses show the $\Delta\chi^2=1$ contours associated with the
   experimental uncertainty.
   The shaded areas represent the
   theory uncertainties due to the variation of additional
   theoretical parameters as described in the text.
 }
 \label{fig:scatterplot}
\end{figure}

The results of the fits are given in Fig.~\ref{fig:scatterplot} and
Table~\ref{tab:fits}.
Fig.~\ref{fig:scatterplot} shows results for $\as$ and $\alpha_0$: the
points give the central result for each $\zeta(C)$ choice, while the
corresponding shaded areas represent  the envelope of results obtained varying
scales and parameters in the theoretical calculation, i.e.\ our
overall theoretical uncertainty.
Each point is accompanied by the $\Delta\chi^2=1$ ellipse, whose
projection along each of the axes defines the $1\,\sigma$ experimental
uncertainty.
Table~\ref{tab:fits} provides the numerical values of the central
results and overall errors for each $\zeta(C)$ choice, and
additionally includes the $\chi^2$ result from the fit,
Eq.~\eqref{eq:chi2}, divided by the number of degrees of freedom.
\begin{table*}
\centering
\begin{tabular}{cccc}
 \toprule
 Model & $\alpha_s(M_Z^2)$ & $\alpha_0(\mu_I^2)$ & $\chi^2/{\rm d.o.f.}$ \\
 \midrule
$\zeta_{0}$ & $0.1121 \pm 0.0006^{+0.0023}_{-0.0014}$ & $0.53 \pm 
0.01^{+0.07}_{-0.04}$ & $1.076$ \\[3pt]
\midrule
$\zeta_{a,1}\equiv\zeta_{b,1}$ & $0.1142 \pm 
0.0005^{+0.0026}_{-0.0015}$ & $0.52 \pm 0.01^{+0.06}_{-0.04}$ & 
$1.045$ \\[3pt]
$\zeta_{a,2}$ & $0.1121 \pm 0.0006^{+0.0024}_{-0.0015}$ & $0.52 \pm 
0.01^{+0.07}_{-0.04}$ & $1.033$ \\[3pt]
$\zeta_{a,3}$ & $0.1099 \pm 0.0007^{+0.0022}_{-0.0014}$ & $0.54 \pm 
0.01^{+0.07}_{-0.05}$ & $1.116$ \\[3pt]
\midrule
$\zeta_{b,2}$ & $0.1163 \pm 0.0005^{+0.0028}_{-0.0017}$ & $0.51 \pm 
0.01^{+0.06}_{-0.04}$ & $1.079$ \\[3pt]
$\zeta_{b,3}$ & $0.1167 \pm 0.0004^{+0.0028}_{-0.0018}$ & $0.53 \pm 
0.01^{+0.06}_{-0.04}$ & $1.143$ \\[3pt]
\midrule
$\zeta_{c}$ & $0.1156 \pm 0.0005^{+0.0027}_{-0.0016}$ & $0.48 \pm 
0.01^{+0.05}_{-0.03}$ & $1.074$ \\[3pt]
 \bottomrule
\end{tabular}
\caption{Results of fits for $\as$ and $\alpha_0$ using the
  different functional forms for 
  $\zeta(C)$ reported in Eq.~\eqref{eq:zeta-forms}.
  The quoted uncertainties encode the total (statistical and
  systematic) experimental 
  uncertainty (first number) and the total theoretical uncertainty
  (second number) estimated as
  described in the text.
  The $\chi^2$ values are those obtained with central scales and
  setup. 
  The results have been obtained with the minimum overlap model,
  Eq.~(\ref{eq:minoverlap}), for correlations between experimental
  systematic uncertainties.  }
    \label{tab:fits}
\end{table*}

The results with the $\zeta_0$ model correspond to the
standard implementation of the leading non-perturbative correction,
which is assumed to amount to a constant shift across the whole $C$
spectrum. The fit returns
\begin{equation}
\zeta_0:~~\alpha_s(M_Z^2) = 0.1121^{+0.0024}_{-0.0016} \,,\quad
\alpha_0(\mu_I^2) = 0.53^{+0.07}_{-0.05}\,,\notag
\end{equation}
and agrees well with that of Ref.~\cite{Hoang:2015hka}, albeit with
larger uncertainties, in part due to our use of NNLL+NNLO rather than
N$^3$LL+NNLO theory predictions.
We observe that several models lead to a $\chi^2$ value that is the
same as, or smaller than, that for the $\zeta_0$ shape. In particular, the
$\zeta_{b,2}$ model returns
\begin{equation}
\zeta_{b,2}:~~\alpha_s(M_Z^2) = 0.1163^{+0.0028}_{-0.0018} \,,\quad \alpha_0(\mu_I^2) = 0.51^{+0.06}_{-0.04}\,,\notag
\end{equation}
with a $\chi^2$ that is similar to that of the
$\zeta_0$ fit. This corresponds to an increase in
$\alpha_s(M_Z^2)$ of about $3.7\%$.
In a number of models ($\zeta_{a,1}\equiv \zeta_{b,1}$, $\zeta_{b,2}$, $\zeta_{b,3}$,
and $\zeta_{c}$) the values of $\alpha_s$ become compatible with the
world average~\cite{PDGQCD2019}
$\alpha_s^{\rm W.A.} = 0.1179 \pm 0.0010$.
The result with the smallest $\chi^2$ is the $\zeta_{a,2}$ model,
which yields a rather small value of
$\as = 0.1121^{+0.0024}_{-0.0016}$.
However the investigations of Appendix~\ref{app:fixed-order}, with a
variety of concrete recoil-scheme prescriptions, seem to disfavour the
$\zeta_{a,2}$ shape, suggesting that yet other factors may be relevant
for maximising the fit quality.

Overall, the results suggest that one should allow for a $3{-}4\%$
uncertainty in $\alpha_s$ extractions from $e^+e^-$ $C$-parameter
data, associated with limitations in our current ability to estimate
hadronisation corrections.

\section{Conclusions}
\label{sec:conclusions}

In this letter we have pointed out that the presence of a Sudakov
shoulder in the differential distribution of some event-shape
observables, such as the $C$ parameter, can be exploited to gain
insight on the observable dependence of the leading ($\sim 1/Q$)
hadronisation correction to the spectrum.
We found that the leading hadronisation correction at the Sudakov
shoulder ($C=3/4$) is over a factor of two smaller than the
corresponding value in the two-jet ($C=0$) limit.

In order to assess the impact of this observation on the fit of the
strong coupling constant, we performed a set of fits using
different assumptions on the scaling of the non-perturbative
correction between the two points. 

Our study is by no means exhaustive, and the inclusion of additional
physical effects (such as the impact of bottom-mass effects) as well
as a careful assessment of other sources of systematic uncertainty
(such as the dependence on the fit range and the choice of correlation
model) is necessary.
However, it clearly reveals that current uncertainties in the
modelling of hadronisation corrections can arguably impact the
extractions of the strong coupling from event shapes at the several
percent level.
In particular, some of the models tested here lead to an increase in
the extracted value of the strong coupling by $3\%-4\%$, which then
becomes compatible with the world average to within uncertainties.
This necessarily raises the question of whether such observables
should still be adopted for percent-accurate determinations of the
strong coupling at LEP energies.
Similar considerations may apply to extractions of $\alpha_s$ obtained
with jet observables, for instance those relying on accurate
calculations for jet
rates~\cite{GehrmannDeRidder:2008ug,Weinzierl:2008iv,Weinzierl:2009ms,DelDuca:2016ily,Banfi:2016zlc}
(e.g. the fits of Refs.~\cite{Dissertori:2009qa,Verbytskyi:2019zhh})
or modifications of $e^+e^-$ event shapes by means of grooming
techniques~\cite{Frye:2016aiz,Baron:2018nfz,Kardos:2018kth} (an
example being the analysis of Ref.~\cite{Marzani:2019evv}).
Further studies are certainly warranted to investigate whether it is
possible to better understand hadronisation for such observables across
their whole spectrum, for example exploiting the large-$n_f$
calculational methods of Ref.~\cite{FerrarioRavasio:2018ubr}.

\section*{Acknowledgements}

One of us (GPS) wishes to thank Zolt\'an Tr\'ocs\'anyi for collaboration in
the early stages of this work.
We are grateful also to Mrinal Dasgupta, Silvia Ferrario Ravasio and
Paolo Nason for numerous discussions on
hadronisation corrections beyond Born configurations.
PM was partly supported by the Marie Sk\l{}odowska Curie
Individual Fellowship contract number 702610 Resummation4PS in the
initial stages of this work.
GPS's work is supported by a Royal Society Research Professorship
(RP$\backslash$R1$\backslash$180112), by the European Research Council
(ERC) under the European Union’s Horizon 2020 research and innovation
programme (grant agreement No.\ 788223, PanScales) and by the Science
and Technology Facilities Council (STFC) under grant ST/T000864/1.

\appendix


\section{Some relevant quantities}
\label{app:formulae}

In the present section we report the expressions for the anomalous
dimensions used in the main text.  The QCD $\beta$ function is defined
by the renormalisation group equation for the QCD coupling constant
\begin{align}
\label{rgealpha}
\frac{\d\alpha_{s}(\mu)}{\d\ln\mu^{2}}=-\alpha_{s}(\mu)\bigg(\frac{\alpha_{s}(\mu)}{\pi}\beta_{0}+\frac{\alpha_{s}^{2}(\mu)}{\pi^{2}}\beta_{1}+\ldots \bigg),
\end{align}
where the first two coefficients read
\begin{align}
\beta_{0} =& \frac{11}{12}C_{A}-\frac{1}{3}T_{F}n_{F},\notag\\
\beta_{1} =& \frac{17}{24}C_{A}^{2}-\frac{5}{12}C_{A}T_{F}n_{F}-\frac{1}{4}C_{F}T_{F}n_{F}.
\end{align}
The $K^{(i)}$ coefficients that appear in the non-perturbative
shift~\eqref{eq:deltaC-master} arise from the perturbative relation
between the strong coupling in the soft physical
scheme~\cite{Catani:1990rr,Banfi:2018mcq,Catani:2019rvy}, denoted here
by $\tilde{\alpha}_s$, and the $\overline{\rm MS}$ coupling
$\alpha_s\equiv \alpha_s(\mu^2)$
\begin{equation}
\label{eq:phys-coupling}
  \tilde{\alpha}_s(\mu^2) = \alpha_s\left(1+\frac{\alpha_s}{2\pi} K^{(1)}
    +\left(\frac{\alpha_s}{2\pi}\right)^2 K^{(2)} + {\cal  O}(\alpha_s^3)  \right)\,.
\end{equation}
They read~\cite{Catani:1990rr,Banfi:2018mcq,Catani:2019rvy}
\begin{subequations}
\begin{align}
 K^{(1)} &= C_A
 \left(\frac{67}{18}-\frac{\pi^2}{6}\right)-\frac{5}{9}n_f\,,
 \\
 K^{(2)} &= C_A^2 \left( \frac{245}{24} - \frac{67}{9}\zeta_2
 + \frac{11}{6}\zeta_3 + \frac{11}{5}\zeta_2^2\right) \notag\\
&+ C_F n_f \left(-\frac{55}{24} + 2\zeta_3\right) \notag\\
& + C_A n_f \left(-\frac{209}{108} + \frac{10}{9}\zeta_2 - \frac{7}{3} \zeta_3\right) 
 - \frac{1}{27} n_f^2 \notag\\
&+ \frac{\beta_0}{2} \left(C_A\left(\frac{808}{27}-28\zeta_{3}\right)-\frac{224}{54}n_f\right)\,.
\end{align}
\end{subequations}

\section{Fixed-order prediction for $\zeta(C)$ and recoil scheme dependence}
\label{app:fixed-order}
It is instructive to repeat the calculation~\eqref{eq:zetaC-def}
starting from a generic $q\bar{q}g$ configuration in the region
$0\! <\! C\! <\! 3/4$. In this case the value of $\Delta C$ in
Eq.~\eqref{eq:deltaC-def} will depend on the specific scheme used to
distribute the recoil due to the emission of the gluer among the
remaining three partons. Therefore, the definition of $\zeta(C)$ away
from the singular points at $C=0$ and $C=3/4$ must
be modified as follows
\begin{align}
  \label{eq:zetaC-nonsing-def}
  \zeta(C) =&\frac{1}{\cal N}
  \lim_{\epsilon\to0}
  \frac{\pi Q}{2\as C_F}
  \int d\Phi_{q\bar{q}g} \,[dk] \, M_{q\bar{q}g}^2(\{p_i\})\,M^2(k)\nonumber\\ \times&
  \Delta C (\{p'_i\},\{p_i\}; k)\delta\left(C-C(\{p_i\})\right)\,
  \delta(k_t - \epsilon)\,,
\end{align}
with the normalisation $\cal{N}$ given by
\begin{equation}
{\cal N}=\int d\Phi_{q\bar{q}g} \,M_{q\bar{q}g}^2(\{p_i\}) \,\delta\left(C-C(\{p_i\})\right)\,.
\end{equation}
In the above two equations $\Phi_{q\bar{q}g}$ denotes the phase space
of the $q\bar{q}g$ system and $M_{q\bar{q}g}^2(\{p_i\})$ is the
corresponding squared amplitude evaluated with the unrecoiled momenta
$\{p_i\}=\{p_q,p_{\bar q},p_g\}$ prior to the emission of the gluer
$k$.
We see that as we approach one of the singular points
$\Delta C (\{p'_i\},\{p_i\}; k)\simeq \Delta C (k)$ and we reproduce
Eq.~\eqref{eq:zetaC-def}.
Away from those points,  the recoil will induce a linear dependence on the
gluer's momentum, hence affecting the value of $\zeta(C)$ in a way
that potentially depends on the specific model of recoil.
The fixed-order calculation of Eq.~(\ref{eq:zetaC-nonsing-def}) will
provide some level of insight into how the leading non-perturbative
correction varies across the spectrum of the event shape.

\begin{figure}
 \includegraphics[width=0.96\linewidth]{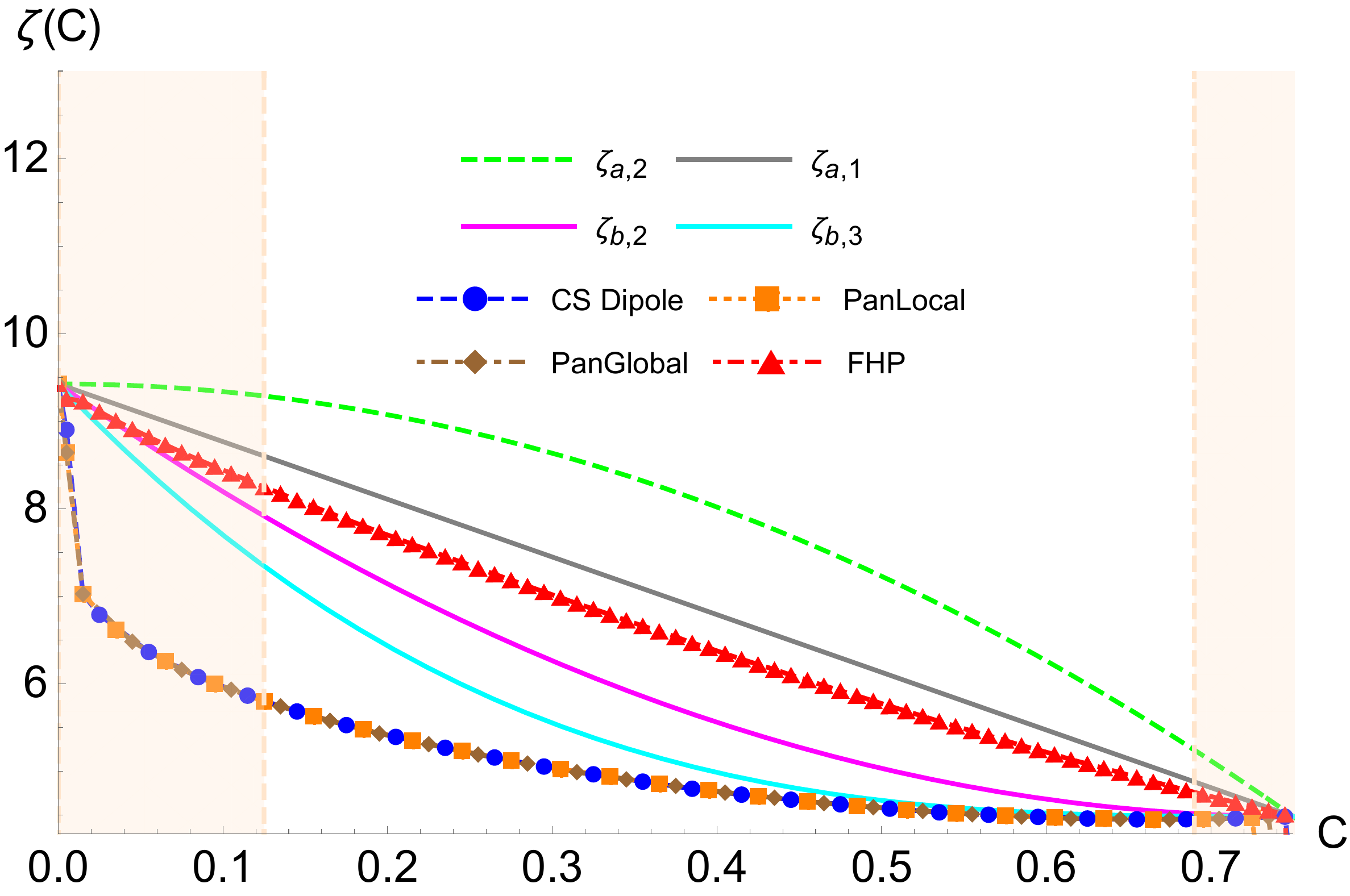}
 \caption{Fixed order calculation of $\zeta(C)$ within different
   recoil schemes, compared to the analytic profiles given in the main
   text. The unshaded area indicates the typical $C$ region where the
   fit of $\alpha_s$ is performed.
   The CS Dipole, PanLocal and PanGlobal results coincide.
 }
 \label{fig:fixed-order}
\end{figure}

For an emission off a given dipole $\{i j\}$ ($\{qg\}$,
$\{q\bar q\}$ or $\{\bar{q} g\}$), we express the gluer's momentum $k$
by means of the Sudakov parametrisation
\begin{equation}
k = \alpha_k p_i + \beta_k p_j + k_{\perp}\,,
\end{equation}
where $\alpha_k=(p_j\cdot k)/(p_i\cdot p_j)$,
$\beta_k=(p_i\cdot k)/(p_i\cdot p_j)$, and
$k_{\perp}=k_t\left[{n}_{\perp,1}\cos\phi+{n}_{\perp,2}\sin\phi\right]$,
with ${n}_{\perp,m}^2 = -1$, ${n}_{\perp,m}\cdot p_{i/j}
= 0$ ($m=1,2$),
${n}_{\perp,1}\cdot{n}_{\perp,2} = 0$.
We consider the following four recoil schemes
\begin{enumerate}
\item {\it CS Dipole}: the scheme is inspired by the Catani-Seymour
  map~\cite{Catani:1996vz}. For an emission $k$ off a dipole $\{ij\}$
  one identifies the emitter and spectator by considering the
  following quantity
 \begin{equation}
   \label{eq:ylk}
   y_{\ell k} = (p_\ell\cdot k)/E_\ell\,,
 \end{equation}
 computed in the event centre-of-mass frame with $\ell=i,j$. The
 emitter is then the dipole end corresponding to the smaller $y_{\ell k}$.
 Once the emitter (say $p_i$) and the spectator ($p_j$) are
 identified, the recoil is distributed as follows
\begin{align}
p^\prime_{i} & =p_{i} - k + \frac{(k\cdot p_i)}{(p_i\cdot p_j)-(k\cdot
               p_j)} p_j\,,\notag\\
p^\prime_{j} & = \left(1-\frac{(k\cdot p_i)}{(p_i\cdot p_j)-(k\cdot
               p_j)}\right) p_j\,.
\end{align}
We also examined an alternative scheme in which the distance
$y_{\ell k}$ is computed in the dipole centre-of-mass frame. The two
schemes produce identical results for the calculation considered in
this appendix, and therefore we omit further discussion of the latter
variant.
\item {\it PanLocal}~\cite{Dasgupta:2020fwr} (antenna variant): the recoil is
    shared locally within the dipole ends as
 \begin{align}
   \label{eq:panlocal}
    p^\prime_{i} & =\alpha_{i}p_{i}+\beta_{i}p_{j}-f k_{\perp}\,,\notag\\
    p^\prime_{j} & =\alpha_{j}p_{i}+\beta_{j}p_{j}-(1-f)k_{\perp}\,.
  \end{align}
  The quantities $\alpha_{i/j}$ and $\beta_{i/j}$ in
  Eq.~\eqref{eq:panlocal} are fully specified by the requirements
  $(p^\prime)^2_{i/j}= 0$, $(p^\prime_i+p^\prime_j+k)=(p_i+p_j)$ and
  $p^\prime_i = p_i$ for $k_t \to 0$. The
  PanLocal~\cite{Dasgupta:2020fwr} rapidity-like variable $\bar\eta$ is
  defined as
  \begin{equation}
    \bar\eta = \ln\left(\frac{\alpha_k}{ k_t}\sqrt{\frac{s_i s_{ij}}{s_j}}\right)\,,
  \end{equation}
  where $s_{i j} =2 p_{i}\cdot p_{j}$, $s_{i}=2 p_{i} \cdot Q$, and $Q$
  is the total event momentum. In the event centre-of-mass frame,
  $\bar \eta=0$ corresponds to a direction equidistant in angle from
  $p_i$ and $p_j$.
  The function $f$ is responsible for sharing the transverse recoil
  among $p_i$ and $p_j$ and it is defined as
  \begin{equation}
    \label{eq:fofeta}
    f \equiv f(\bar\eta) = \frac{e^{2\bar\eta}}{1+e^{2\bar\eta}}\,.
  \end{equation}
  Finally, we have
  \begin{align}
    \alpha_{i} & =\frac{(\sqrt{\lambda_{1}}+\sqrt{\lambda_{2}})^{2}+4f^{2}}{4(1-\beta_k)}\,\alpha_k\beta_k\,,\notag\\
    \beta_{i} & =\frac{(\sqrt{\lambda_{1}}-\sqrt{\lambda_{2}})^{2}+4f^{2}}{4(1-\alpha_k)}\,\alpha_k\beta_k\,,\notag\\
    \alpha_{j} & =\frac{(\sqrt{\lambda_{1}}-\sqrt{\lambda_{2}})^{2}+4(1-f)^{2}}{4(1-\beta_k)}\,\alpha_k\beta_k\,,\notag\\
    \beta_{j} & =\frac{(\sqrt{\lambda_{1}}+\sqrt{\lambda_{2}})^{2}+4(1-f)^{2}}{4(1-\alpha_k)}\,\alpha_k\beta_k\,,
  \end{align}
  with
  \begin{align}
    \label{eq:IIb-lambda}
    \lambda_{1}  &=(1-\alpha_k-\beta_k)/(\alpha_k \beta_k)\,,\notag\\
    \lambda_{2}  &= \lambda_{1}  + 4 f(1-f)\,.
  \end{align}

\item {\it PanGlobal}~\cite{Dasgupta:2020fwr}: the longitudinal
  recoil is assigned locally within the dipole as
  \begin{align}
    \bar p_{i} & =(1-\alpha_{k})p_{i}\,,\notag\\
    \bar p_{j} & =(1-\beta_{k})p_{j}\,,
  \end{align}
  and the transverse recoil is assigned by applying a Lorentz boost
  and a rescaling to the full event so as to obtain final momenta
  $\{p^\prime,k^\prime\}$ whose sum gives the original total momentum
  $Q$ (see~\cite{Dasgupta:2020fwr} for details).\\

\item {\it FHP}: the scheme is inspired by that proposed by
  Forshaw-Holguin-Pl\"atzer in Ref.~\cite{Forshaw:2020wrq}. It is
  similar to PanGlobal, with the difference that only the longitudinal
  recoil along the emitter, say $p_i$, is assigned locally
  \begin{align}
    \bar p_{i} & =(1-\alpha_{k})p_{i}\,,
  \end{align}
  and the remaining longitudinal and transverse recoil is assigned by
  applying a Lorentz boost and a rescaling to the full event as in the
  PanGlobal scheme. Unlike the proposal in the original
  paper~\cite{Forshaw:2020wrq}, we identify the emitter $p_i$ with the
  dipole end closer in angle to $k$ in the event centre-of-mass frame,
  that is the one with the smaller $y_{\ell k}$ defined in
  Eq.~\eqref{eq:ylk}.
 For our purposes, this is physically similar to what is done in
 Ref.~\cite{Forshaw:2020wrq}.
\end{enumerate}
The results of the computation are reported in
Figure~\ref{fig:fixed-order}, where for comparison we also report the
curves corresponding to the profiles $\zeta_{a,1}\equiv
\zeta_{b,1}$, $\zeta_{a,2}$ and $\zeta_{b,2}$.
We observe that the CS Dipole, PanLocal and PanGlobal schemes yield
nearly identical results for $\zeta(C)$, which depart very sharply
from the asymptotic value in the two-jet limit and approach the shape
of the $\zeta_{b}$-type profiles at large values of $C$.
Instead, the FHP scheme gives a less convex shape, close to a linear
scaling in the fit region (indicated by the unshaded area in the
plot).

We believe that the similarity between the CS Dipole, PanLocal and
PanGlobal schemes originates from the fact that, in the presence of a
single perturbative gluon, $\zeta(C)$ is largely insensitive to the
precise distribution of the transverse recoil among the particles in
the event, which at this order and for this particular observable, is
washed out by the integration over the azimuth and rapidity of the
perturbative gluon $p_g$.
Conversely, the result does seem to depend on how the longitudinal
recoil is assigned.
The CS Dipole, PanLocal and PanGlobal schemes schemes all assign
the longitudinal recoil locally within the emitting dipole, while in
the FHP scheme part of the longitudinal recoil is shared among all
particles in the event.

Note that all recoil schemes appear to be below the $\zeta_{a,1}\equiv\zeta_{b,1}$
model.
This tends to disfavour the $\zeta_{a,2}$ type model for interpolation
of $\zeta(C)$ between $C=0$ and $C=3/4$, even though it gave the
lowest $\chi^2$ in the $\alpha_s$ fits in Section~\ref{sec:fit}.

A final comment concerns the limitations of the fixed-order nature of
the study carried out in this appendix.
At the order at which we work, the fact that we force the perturbative
$q\bar q g$ system to have a given value of the $C$-parameter causes
the perturbative gluon to have a hardness comparable to $C Q$.
Were we to go to higher orders, it would become possible for the
perturbative event to contain additional, much softer gluons.
Those gluons could also be involved in the recoil from the
non-perturbative gluer, further altering the gluer's impact on the
$C$-parameter.
To take this into account, one would need to carry out a non-global
type resummation that includes any number of perturbative soft gluons
between the momentum scale set by the $C$-parameter value and the
non-perturbative scale.
First investigations in this direction (with just the PanGlobal recoil
for the gluer) suggest that the impact of the resummation is
non-negligible, and also continue to favour $\zeta(C)$ profiles that
are below the linear $\zeta_{a,1}\equiv\zeta_{b,1}$ profile.
\logbook{a464fc7f3}{see ../cpar-panscales/results/cpar-panscales.pdf}

\section{Fits with uncorrelated systematic experimental uncertainties}
\label{app:uncorr}

In this appendix we report the simultaneous fit of $\alpha_s$ and
$\alpha_0$ obtained with the same procedure outlined in the main text,
albeit replacing the model~\eqref{eq:minoverlap} with the simpler
assumption of uncorrelated systematic uncertainties in the
experimental data, namely
\begin{equation} 
  E_{ij} = \delta\sigma_{{\rm
        syst},i}^2\,\delta_{ij}\,.
\end{equation}
The results are given in Table~\ref{tab:fits_uncorr}.
Relative to Table~\ref{tab:fits}, the absence of correlations leads to
an increase in the experimental uncertainties for $\as$ of
${\cal O}(10\%-20\%)$, and the $\chi^2$ values decrease.
The central $\as$ and $\alpha_0$ results are essentially unchanged, as
are the theory systematics.

\begin{table*}
\centering
\begin{tabular}{cccc}
 \toprule
 Model & $\alpha_s(M_Z^2)$ & $\alpha_0(\mu_I^2)$ & $\chi^2/{\rm d.o.f.}$ \\
 \midrule
$\zeta_{0}$ & $0.1122 \pm 0.0007^{+0.0024}_{-0.0014}$ & $0.52 \pm 
0.01^{+0.07}_{-0.04}$ & $0.813$ \\[3pt]
\midrule
$\zeta_{a,1}\equiv\zeta_{b,1}$ & $0.1142 \pm 
0.0006^{+0.0026}_{-0.0015}$ & $0.52 \pm 0.01^{+0.06}_{-0.04}$ & 
$0.796$ \\[3pt]
$\zeta_{a,2}$ & $0.1121 \pm 0.0007^{+0.0024}_{-0.0015}$ & $0.52 \pm 
0.01^{+0.07}_{-0.04}$ & $0.787$ \\[3pt]
$\zeta_{a,3}$ & $0.1100 \pm 0.0008^{+0.0022}_{-0.0014}$ & $0.54 \pm 
0.01^{+0.07}_{-0.05}$ & $0.845$ \\[3pt]
\midrule
$\zeta_{b,2}$ & $0.1162 \pm 0.0005^{+0.0028}_{-0.0017}$ & $0.51 \pm 
0.01^{+0.06}_{-0.04}$ & $0.822$ \\[3pt]
$\zeta_{b,3}$ & $0.1167 \pm 0.0005^{+0.0028}_{-0.0018}$ & $0.53 \pm 
0.01^{+0.06}_{-0.04}$ & $0.870$ \\[3pt]
\midrule
$\zeta_{c}$ & $0.1156 \pm 0.0006^{+0.0027}_{-0.0016}$ & $0.48 \pm 
0.01^{+0.05}_{-0.03}$ & $0.807$ \\[3pt]
 \bottomrule
\end{tabular}
\caption{
  Results of fits for $\as$ and $\alpha_0$ using the
  different functional forms for 
  $\zeta(C)$, as in Table~\ref{tab:fits}, but with an assumption of
  uncorrelated experimental uncertainties instead of the minimum
  overlap model.
}
    \label{tab:fits_uncorr}
\end{table*}

\section{Modified logarithms and comparison to profile functions}
\label{app:modlogs}
In order to properly ensure that the resummation is turned off at the
kinematic endpoint of the differential distribution, we modify the
resummed logarithms by making the replacement
\begin{equation}
\label{eq:modlog}
\ln\frac{6 \,x_C}{C}\to L \equiv \frac{1}{p} \ln\left(1 + \frac{\left( 6\,
      x_C\right)^p}{C^p} -
  \frac{\left( 6
      x_C\right)^p}{C_{\rm max}^p}\right)\,,
\end{equation}
where $p$ denotes a positive parameter, and $C_{\rm max}$ is the
kinematic endpoint of the $C$-parameter distribution in the
multi-jet regime, i.e. $C_{\rm max} = 1$. 
The prescription of Eq.~\eqref{eq:modlog} is but a possible choice and
other sensible solutions can be found in the literature (see
e.g. Ref.~\cite{Hoang:2015hka}). This ambiguity introduces an
additional theoretical uncertainty in the calculation that must be
carefully estimated.
The quantity $x_C$ is of order one and its variation estimates the
resummation uncertainty due to missing higher-logarithmic corrections.
Specifically, the $\Sigma^{\rm N^kLL}(C)$ resummed cross section
acquires a net $x_C$ dependence such that,
\begin{equation}
\frac{d\Sigma^{\rm N^kLL}(C)}{d\ln x_C}={\cal O}(\rm{N^{k+1}LL})\,,
\end{equation}
in the logarithmic limit as $C\to 0$.
Similarly, the parameter $p$ determines how quickly the
resummation is turned off in the region $C\sim C_{\rm max}$.

\begin{figure}
 \includegraphics[width=0.96\linewidth]{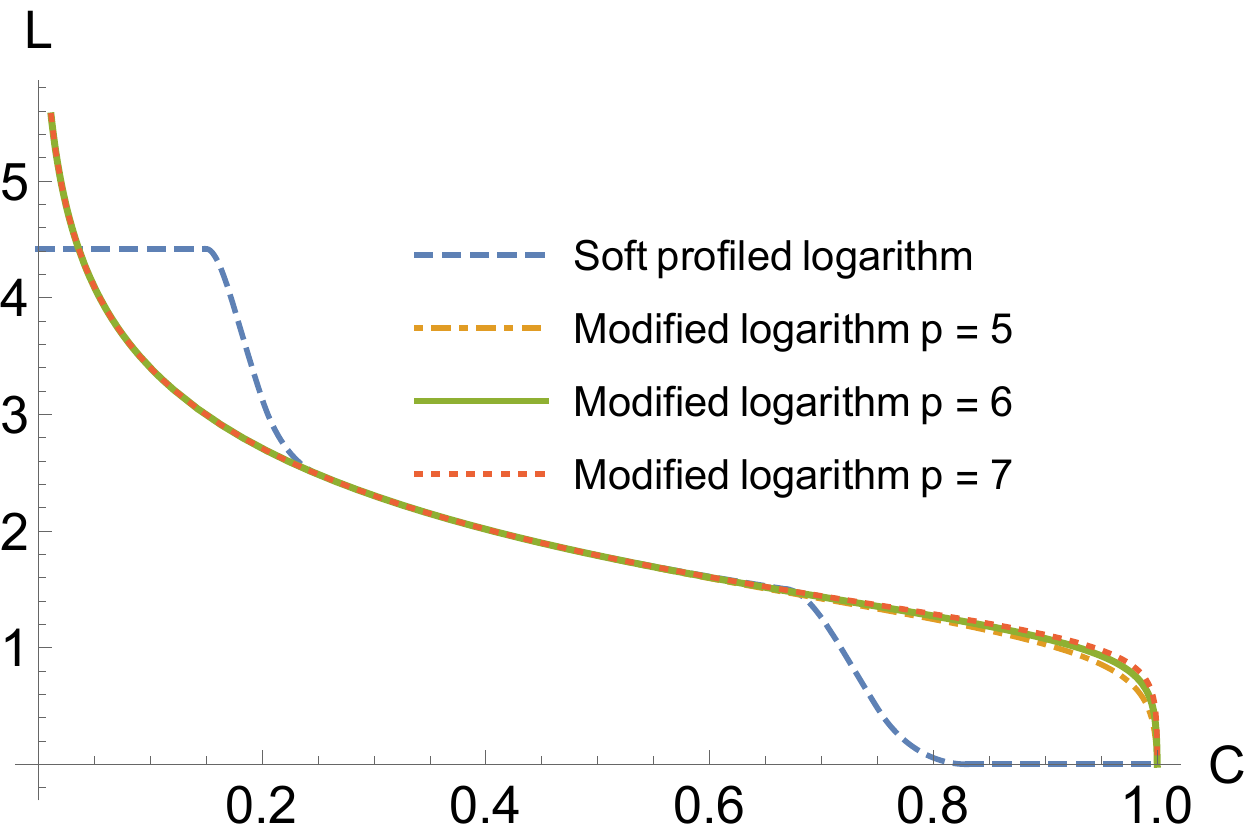}
 \caption{The figure displays a comparison between the resummed
   modified logarithm~\eqref{eq:modlog} with different $p$ values and
   the profiled logarithm $\ln(Q/\mu_s(C))$ where the profiled soft
   scale $\mu_s(C)$ is defined in Ref.~\cite{Hoang:2015hka}. The
   centre-of-mass energy $Q$ is set to the $Z$-boson mass.}
 \label{fig:profile-logs}
\end{figure}

The choice of $p$ must guarantee that the resummation does not
substantially affect
the prediction in regions of the spectrum dominated by hard
radiation.
An inspection of the first-order $C$-parameter distribution reveals
contributions suppressed by a (linear) power of $C$ relative to the
dominant $(\ln C)/C$ dependence.
Were we to take $p=1$, the first-order expansion of the resummation
would be associated with perturbative linear power-suppressed
contributions whose coefficient would be larger than that
observed in the exact fixed-order calculation.
Accordingly, we believe it is sensible to apply the restriction $p>1$
to avoid such contributions, and ensure that the resummation does not
affect the dominant scaling at subleading power.
\logbook{7b0ce3cb}{see ../subleading_power_studies/expanded_minus_LO_difference.pdf}%
With this constraint, we find that the extracted value of $\alpha_s$
depends only very mildly on the choice of $p$ and well within the
quoted theoretical uncertainties.
We then choose $p = 6$ and $x_C=1/2$ and vary both parameters as
outlined in Section~\ref{sec:fit} in the uncertainty estimate.
This specific choice is motivated by the fact that the scaling of the
modified logarithm~\eqref{eq:modlog} in most of the fit range that we
adopted in Section~\ref{sec:fit} happens to reproduce that of the
profiled logarithms of the soft function of Ref.~\cite{Hoang:2015hka},
which we use as a reference benchmark in our study. A comparison
between the two prescriptions is shown in
Figure~\ref{fig:profile-logs}

\bibliographystyle{spphys}
\bibliography{biblio}

\end{document}